\documentclass{article}
\usepackage{amsmath,amssymb,amsthm,graphicx,hyperref}
\usepackage{authblk}
\usepackage[margin=2cm]{geometry}
\usepackage[utf8]{inputenc}
\DeclareUrlCommand\email{\urlstyle{rm}}
\begin{document}
	\title{Relativistic Gas: Lorentz-invariant Distribution for the velocities}
	\author{$^{a,b}$Evaldo M. F. Curado}
	\author{$^a$Carlos E. Cede\~no} %
	\author{$^a$Ivano Dami\~ao-Soares}%
	\author{$^{a,b,c,d}$Constantino Tsallis}%
	\affil{$^a$Centro Brasileiro de Pesquisas Fisicas, Rua Xavier Sigaud 150, Rio de Janeiro 22290-180, Brazil}
	\affil{$^b$National Institute of Science and Technology for Complex Systems, Rua Xavier Sigaud 150, Rio de Janeiro 22290-180, Brazil}
	\affil{$^c$Santa Fe Institute, 1399 Hyde Park Road, Santa Fe, New Mexico 87501, USA}
	\affil{$^d$Complexity Science Hub Vienna, Josefst\"adter Strasse 39, 1080 Vienna, Austria}
	\date{\today}
	\maketitle
	\email{evaldo@cbpf.br}
	\email{cecedeno@cbpf.br}
	\email{ivano@cbpf.br}
	\email{tsallis@cbpf.br}
\begin{abstract}
In 1911, J\"uttner proposed the generalization, for a relativistic gas, of the Maxwell-Boltzmann distribution of velocities. Here we want to discuss, among others,  J\"uttner probability density function (PDF). Both the velocity space and, consequently, the momentum space are not flat in special relativity. The velocity space corresponds to the Lobachevsky one, which has a negative curvature. This curvature induces a specific power for the Lorentz factor in the PDF, affecting J\"uttner normalization constant in one, two, and three dimensions. Furthermore, J\"uttner distribution, written in terms of  a more convenient variable, the rapidity, presents a curvature change at the origin at sufficiently high energy, which does not agree with our computational dynamics simulations of a relativistic gas. However, in one dimension, the rapidity satisfies a simple additivity law. This allows us to obtain, through the Central Limit Theorem, a new, Lorentz-invariant, PDF whose curvature at the origin does not change for any energy value and which agrees with our computational dynamics simulations data. Also, we perform extensive first-principle simulations of a one-dimensional relativistic gas constituted by light and heavy particles. 
 
\end{abstract}

\maketitle

\begin{quotation}
The proper relativistic generalization of the Maxwell distribution of velocities of an ideal gas is still an open question. The main proposal, Jüttner distribution, was introduced in 1911 and many attempts to improve it do exist. We show here that neither the Jüttner nor the so-called Modified Jüttner  distributions are Lorentz-invariant. The Jüttner distribution, in particular, was constructed on the basis of a flat velocity space whereas this space is curved, namely the Einstein-Lobachevsky space.  In the one-dimensional case, by using the relativistic addition of velocities and the central limit theorem, we were able to build a Lorentz-invariant distribution (LID). We performed molecular dynamical simulations of a one-dimensional relativistic gas and verified that, at high enough energy, the Jüttner and the Modified Jüttner distributions do not match the numerical simulations, in contrast with the LID which does agree. Additionally, we exhibit how the corresponding temperatures of these three distributions behave as functions of the gas mean energy $\langle E\rangle$. An important corollary of our framework is that the temperature of the relativistic gas is a Lorentz-invariant quantity. 	 
\end{quotation}

\section{Introduction}
\par Probability density functions (PDF) for the velocity of the particles of a  relativistic gas are, actually, controversial themes in both astrophysics \cite{Sadegzadeh18,Molnar20,Melrose21} and particle physics of heavy ions\cite{Adcox01,Adler02,Marques15}. 
The first proposal and one of the most well-known probability distribution of velocities for a relativistic gas was originally derived by J\"uttner \cite{Juttner} in 1911. Until now, it remains as an interesting attempt to construct the basis of relativistic thermodynamics. 

\par J\"uttner constructed his distribution maximizing the entropy $S$ of a relativistic gas, conserving constant the   relativistic energy $E$ (a quantity which is not Lorentz-invariant) and the number $N$ of particles.  The distribution derived by him  in the limit of velocities much smaller than the velocity of the light, $v \ll c$, reproduces the classical Maxwell-Boltzmann  distribution of velocities \cite{boltzmann1872,boltzmann1877,Maxwell_1,Maxwell_2}.  However, the J\"uttner  distribution is {\it not} a Lorentz-invariant one as we will point out soon. Moreover,  this distribution has been, mainly in the last twenty years,  a matter of intense debate  in the literature \cite{Landsberg66,Landsberg67,cubero2007,Montakhab09,Ghodrat11,CGS16,Kubli2021}. 

\par In a previous paper Curado et al. \cite{CGS16} worked on the construction of Lorentz-invariant distributions, where they explored histograms with constant bins in both velocity and rapidity, using the velocity as the abscissa of their histograms. The distributions obtained there are Lorentz-invariant but in terms of the velocity instead of the rapidity.  

\par Here, we extended the work of Curado et al. \cite{CGS16}, showing the rapidity as a natural variable 
to construct Lorentz-invariant distributions, with the aim to describe the statistical behaviour of this kind of systems. We compare the different temperatures predicted for Jüttner, the modified one, and the Lorentz-invariant distribution (LID). We also derive from the maximum entropy principle the proper Lorentz-invariant distribution in terms of the rapidity. Here, we exhibit the non Lorentz-invariant character of the J\"uttner distribution and also an attempt to adapt the J\"uttner distribution to the requirements of the Lorentz transformation,	the so-called  Modified J\"uttner distribution.  We expose the missteps within the J\"uttner derivation and present, for the one-dimensional case, a new Lorentz-invariant distribution (LID) based on the variable rapidity. The latter is defined in a flat space, contrary to the velocity, which is curved. Because of this, the rapidity is an ideal variable for constructing a new Lorentz-invariant distribution, as presented later on.  This novel probability density distribution is derived from first principles. We show that this new distribution fits with high accuracy the numerical data of molecular dynamics simulation for the rapidity distribution of one-dimensional gases. 

\par To present our discussion, we divide this paper as follows. 
Section I is the introduction. In Section 2  we discuss the original work by J\"uttner 
\cite{Juttner} and show why his distribution - and also the so-called modified J\"uttner distribution - are not Lorentz invariant PDF.  In 
Section III we essentially follow the derivation of the new LID presented in \cite{CGS16} with a slightly different approach. In Section IV 
we present new molecular dynamics simulations we have performed, clearly showing that this novel Lorentz-invariant distribution (LID) matches well the numerical simulations for relativistic gases. We also verify that both the J\"uttner and the Modified J\"uttner distributions fail in this respect. 
The analytical expressions of the average energy of each particle were calculated and compared, showing the possible discrepancies in the temperature estimates depending on which distribution is used. In Section V we show how the LID could be derived from a MaxEnt principle. 
Finally, we discuss our findings in the conclusions.

\section{The Jüttner and the modified Jüttner distributions}
In his original work\cite{Juttner}, J\"uttner expressed the number of particles $dN$ of a given gas in a volume element in the phase-space $(x,y,z,p_x,p_y,p_z)$ as $dN=\tilde{P}_J(x,y,z,p_x,p_y,p_z) \, dx\,dy\,dz\,dp_x\,dp_y\,dp_z$, where $p_i$ are the reduced relativistic linear momentum components, or the relativistic  linear momentum per unit mass and $(x, y, z)$ denote the position. 

Thus the total number of particles is given by
\begin{equation}
	\label{eq1}
	N=\int \tilde{P}_J\, dx\,dy\,dz\,dp_x\,dp_y\,dp_z\,,
\end{equation}   
where $p_i=\gamma(v)\dot{x}_i$, $ v_i = \dot{x}_i  \, , \, i = x, y, z$,  $v=\sqrt{\dot{x}+\dot{y}+\dot{z}}$, being $v$ the velocity of a particle of the system with respect to an inertial frame, and $\gamma(v)\equiv(1-v^2/c^2)^{-1/2}$ is the Lorentz factor. The mean energy of the particles is 
\begin{equation}
	\label{eq2}
	E=m_0c^2\int \gamma(v) \tilde{P}_J dx\,dy\,dz\,dp_x\,dp_y\,dp_z\,,
\end{equation}
where $m_0$ is the rest mass, and $c$ is the speed of light. The entropy $S$ of the system associated with such a state is
\begin{equation}
	\label{eq3}
	S=-k\int \tilde{P}_J\ln \tilde{P}_J dx\,dy\,dz\,dp_x\,dp_y\,dp_z.
\end{equation}
Maximizing the entropy $S$, and conserving as constants both $N$ and the relativistic energy $E$, J\"uttner  found a probability distribution of velocities for a relativistic gas in three-dimensions, $\tilde{P}_J(\gamma \, v)$. Transforming the phase space $(x, y, z, p_x, p_y,p_z ) \rightarrow (x, y, z, v_x, v_y, v_z )$, J\"uttner obtained his three-dimensional relativistic distribution of velocities, $P_J(v_x, v_y, v_z)$. This distribution  yields, as expected, zero probability for a particle to have a speed larger than that of light, $c$. 

This three-dimensional distribution can be generalized  to a $d$-dimensional distribution of velocities as follows: 
\begin{subequations}
	\label{eq4}	
	\begin{align}
		\label{eq4a}
		p_J(v)\,d^dv&=\dfrac{m_0^d\gamma^{d+2}(v) e^{- \beta_J \gamma(v)}}{Z_J}d^dv, \nonumber\\
		Z_J&=\begin{cases} 2 m_0 c K_1(\beta_J)\quad &\text{for}\quad d=1 \\
			\dfrac{4 \pi  m_0^2c^2  (\beta_J+1) e^{-\beta_J} }{\beta_J^2} \quad &\text{for}\quad d=2 \\  
			\dfrac{8 \pi  m_0^3 c^3  K_2(\beta_J)}{\beta_J} \quad &\text{for}\quad d=3
		\end{cases}\,,
	\end{align}
\end{subequations}
where $d$ is the dimension, $d^dv=\{dv,dv_x\,dv_y,dv_x\,dv_y\,dv_z\}=\{dv,vdv\,d\theta,v^2\sin\theta dv\,d\theta\,d\phi\}$, for $d = 1, 2, 3$ dimensions respectively, $\beta_J=m_0 c^2/(kT_J)$, $k$ is the Boltzmann constant, $T_J$ is the temperature associated with the J\"uttner distribution, $Z_J$ is the normalization constant, and,  $K_1(x)$ and $K_2(x)$ are the modified Bessel functions of the second kind.

\par It is worthwhile to note that J\"uttner (wrongly) considered a flat space for the velocities (it should be remembered that this distribution was proposed as early as in 1911), as shown above. In fact, the relativistic space of velocities is not flat, it is a curved space, with negative curvature, more precisely a Lobachevsky space. This ensures that the pre-factor before the exponential in the distribution is not $\gamma^{d+2}$, as in the Eq. (\ref{eq4a}), but rather the pre-factor $\gamma^{d+1}$ which, with $d^d v$, is a Lorentz invariant quantity. This is one flaw in the J\"uttner distribution, since $\gamma^{d+2} d^d v$ is not a Lorentz-invariant quantity. A second problem is the argument of the exponential in Eq. (\ref{eq4a}), the relativistic energy. It is well-known that the energy is not a Lorentz-invariant quantity. These two aspects make the J\"uttner distribution a non Lorentz-invariant distribution. 

\par In an attempt to correct J\"uttner's distribution, at least at the pre-factor level, Lehmann \cite{Lehmann06}, and a year later, Dunkel and H\"anggi \cite{Dunkel_Hanggi_07}, modified the J\"uttner's distribution as follows:
\begin{subequations}
	\label{eq5}
	\begin{align}
		\label{eq4b}
		p_{MJ}(v)\,d^dv&=\dfrac{m_0^{d-1}\gamma^{d+1}(v) e^{- \beta_{MJ} \gamma(v)}}{Z_{MJ}}\,d^dv, \nonumber \\
		Z_{MJ}&=\begin{cases}
			2 c K_0( \beta_{MJ})   \quad &\text{for}\quad d=1 \\ 
			\dfrac{4 \pi  m_0 c^2 e^{-\beta_{MJ}} }{\beta_{MJ}} \quad &\text{for}\quad d=2 \\ 
			\dfrac{8 \pi m_0^2 c^3  K_1(\beta_{MJ})}{\beta_{MJ}}
			\quad &\text{for}\quad d=3 \\
		\end{cases}\,,
	\end{align}
\end{subequations}
replacing the factor $\gamma^{d+2}$ by the factor $\gamma^{d+1}$; $\beta_{MJ}\equiv m_0 c^2/(kT_{MJ})$, where $T_{MJ}$ is the temperature associated to this distribution and $Z_{MJ}$ is its normalization constant.

\par It is worth emphasizing that neither the momentum nor the velocity spaces are flat in special relativity and that the relativistic energy is not a Lorentz-invariant quantity. Either for both of these reasons, or for only one of them, neither J\"uttner nor Modified J\"uttner are Lorentz-invariant distributions for a relativistic gas. 

\par
\begin{figure}[h!]
	\begin{center}
		\begin{tabular}{ccc}
			\includegraphics[width=5.5cm]{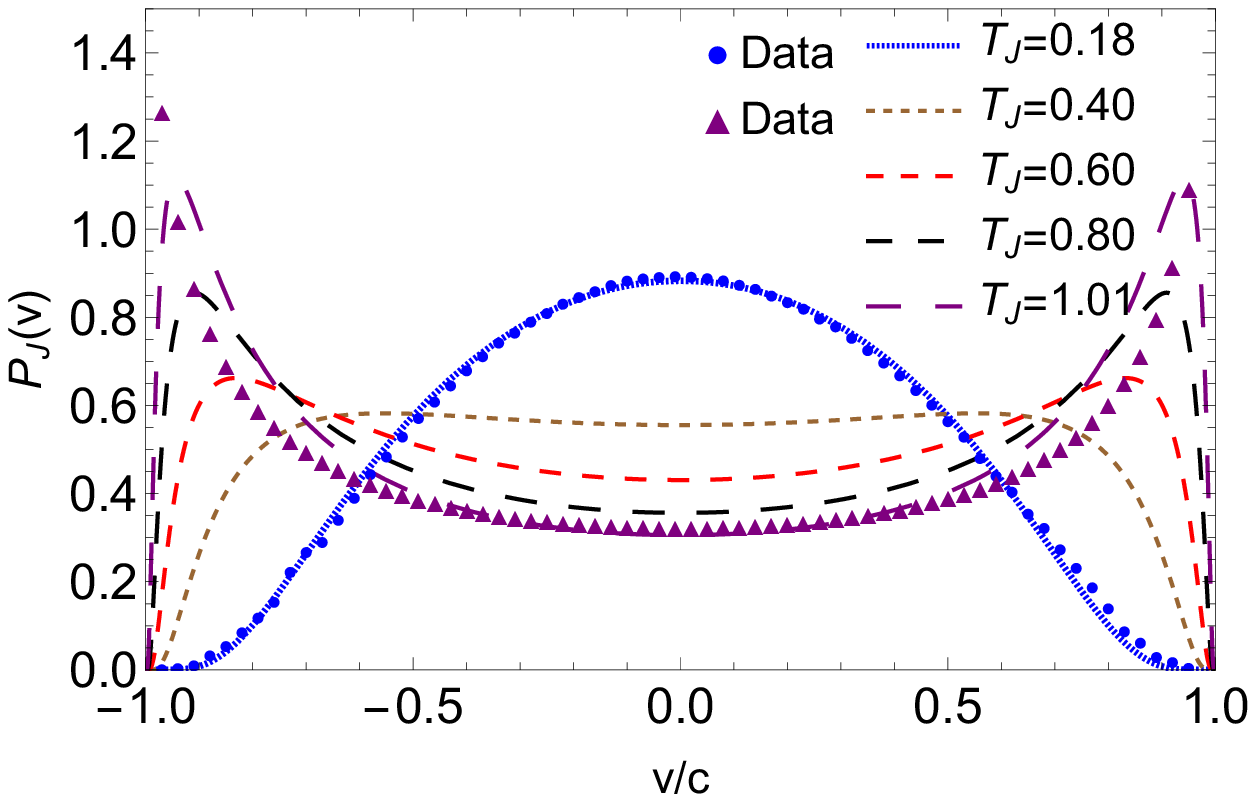} &
			\includegraphics[width=5.5cm]{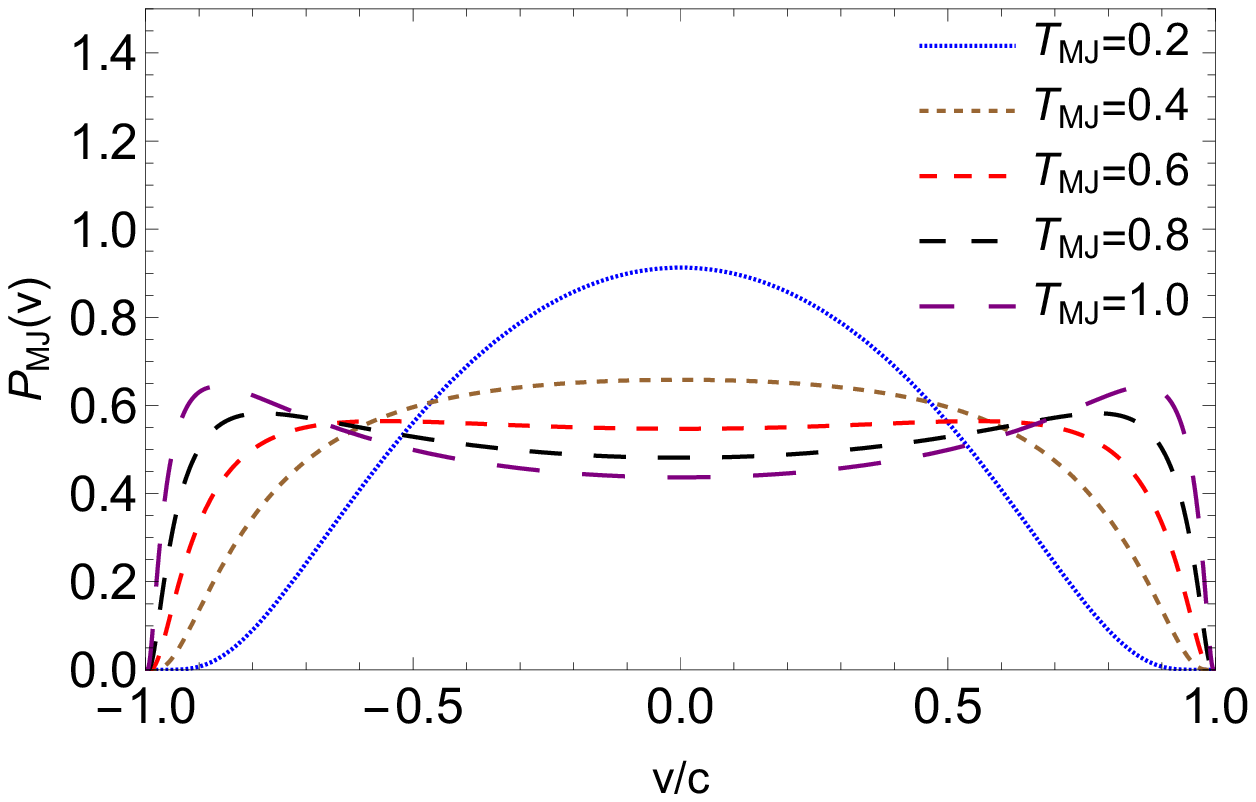} &
			\includegraphics[width=5.5cm]{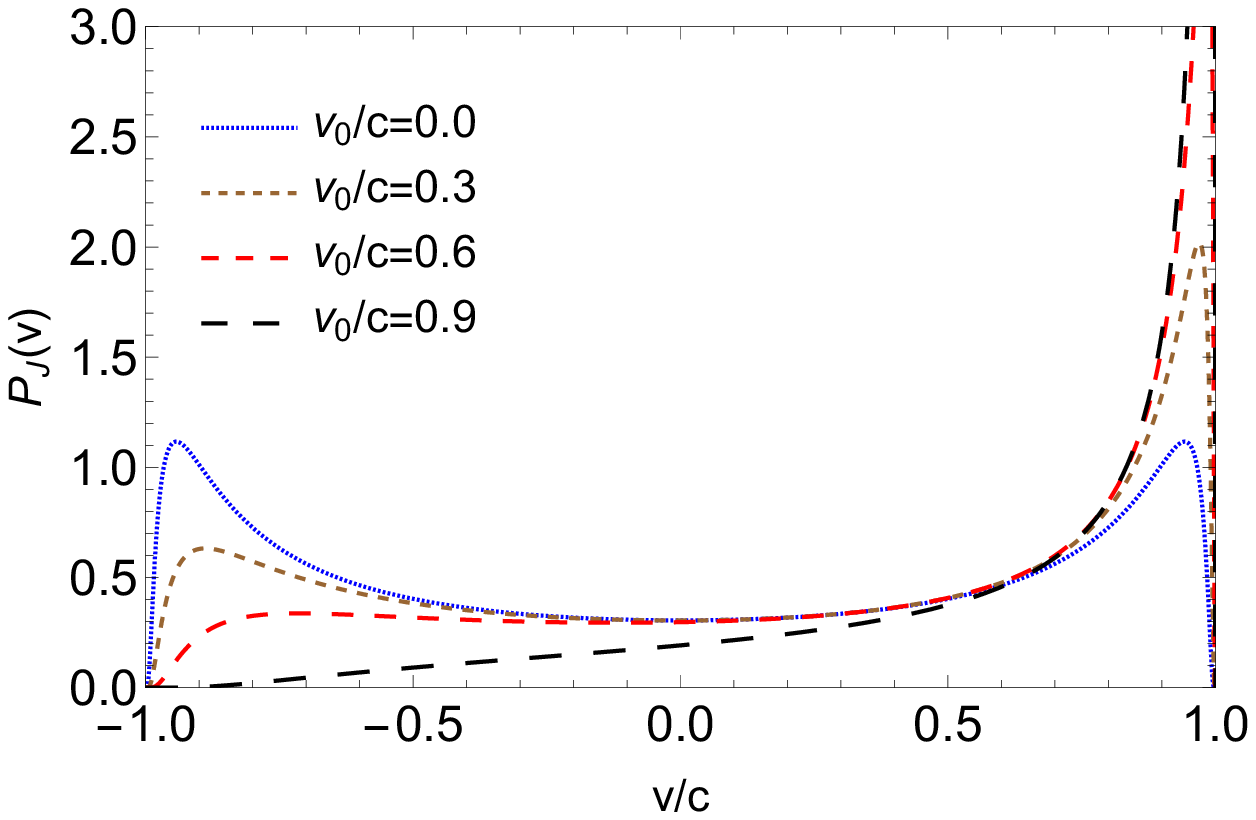}
		\end{tabular}
	\end{center}
	\caption{One-dimensional distributions. Left panel: $P_J(v)$ as a function of the velocity for different values of $T_J$. We plot here the computational results of our simulations, which are explained in details below, for comparison with the analytical result. Middle panel: $P_{MJ}(v)$ as a function of the velocity. We use the same parameters as in the left panel, for comparison. Both functions recover the Maxwell-Boltzmann distribution at the limit $T\rightarrow 0$ and become bi-modal for high values of $T$. Right panel: $P_J(v)$, for $T_J=1.0$, as a function of the velocity as measured by different external observers, see Eq. \eqref{eq9a}, for $v_0/c=0.0,\,,0.3,\,0.6,\,0.9$. We consider in all these plots $m_0=c=k=1$.} 
	\label{fig1}
\end{figure}

 We note that J\"uttner distribution $P_J(v)$ and the Modified J\"uttner distribution $P_{MJ}(v)$ change their behavior from mono-modal to bimodal functions for $T_J = m_0c^2 /((2 + d) k)$ and $T_{MJ} =  m_0 c^2/((1 + d) k)$, respectively. For comparison,  we plot in the left and the middle panels of Figs. (\ref{fig1}) the one dimensional J\"uttner  $P_J(v)$ and the Modified J\"uttner $P_{MJ}(v)$ distributions as functions of the  velocity, respectively. Note that the concavity change in both distributions at the origin occurs at the temperatures $T_{J} = m_0c^2/(3 k)$ and ${T_{MJ} =  m_0c^2/(2 k)}$ respectively. In the low-temperature regime both distributions  reproduce the classical Maxwell-Boltzmann one, when the particles move in the low-velocity limit $v\ll c$. In the figures we use normalised units $c=k=m_0=1$.

\par  Now, we exhibit that the J\"uttner and the Modified J\"uttner distributions for the velocities, in one dimension, are not Lorentz-invariant. It implies that both J\"uttner and the Modified J\"uttner distributions seen by one observer with velocity $v_0$ with respect to the laboratory frame must be distinct from those seen by the laboratory frame.  Using the relative velocity $v^\prime = (v - v_0)/(1-v v_0))$	into $P_{J}(v)$ and $P_{JM}(v)$ we obtain 
\begin{align}
	\label{eq9a}
	P_J(v^\prime)dv^\prime &=
	\dfrac{c^2 \exp\left(-\beta_J \left(1-\frac{c^2 (v-v_{0})^2}{\left(c^2-v v_{0}\right)^2}\right)^{-1/2}\right)}{2 \left(c^2-v^2\right) K_1(\beta_J) \sqrt{1-\frac{c^2
				(v-v_{0})^2}{\left(c^2-v v_{0}\right)^2}}} dv \nonumber \\
		&\ne  p_J(v)dv \,,
			\\
	P_{MJ}(v^\prime)dv^\prime &= \frac{c \exp \left(-\beta_J\left(\frac{\left(c^2-v^2\right) \left(c^2-v_{0}^2\right)}{\left(c^2-v v_{0}\right)^2}\right)^{-1/2}\right)}{2
		\left(c^2-v^2\right) K_0(\beta_J)}
	dv \nonumber \\
&\ne p_{MJ}(v)dv\,.
	\label{eq9b}
\end{align}	
We plot J\"uttner's distribution \eqref{eq9a} under a Lorentz transformation, for $T = 1.0$, in the right panel of Fig. \ref{fig1},  taking normalized units. We verify that for different velocity values of the external observer $v_0$, the distribution form is not maintained invariant under Lorentz transformations.

\section{On a Lorentz invariant distribution}
The  relativistic one-dimensional addition of velocities of two particles, viewed by an observer (at rest with the box) as having velocities $v_1$ and $v_2$, satisfies	the well-known formula,
\begin{equation}
	\label{eqaddition}
	v=\dfrac{v_1 + v_2}{1+v_1v_2/c^2},
\end{equation}
which can be rewritten in a more symmetric way  as
\begin{equation}
	\label{eqaddition2}
	\frac{1+v}{1-v} = \frac{1+v_1}{1-v_1} \frac{1+v_2}{1-v_2} .
\end{equation}
Notice that we are considering here  the velocities $v_1$ and $v_2$ as the {\it the relativistic relative} velocities between particles $1$ or $2$ and the velocity of the reference frame (box), which is taking here equal to zero; $v$ is a relative velocity with respect	to the velocity of the box. As well-known, relative velocities are Lorentz-invariant quantities.

\noindent If we are adding $N$ particles, Eq. (\ref{eqaddition2}) can be easily generalized  as follows:
\begin{equation}
	\label{eqaddition3}
	\frac{1+v}{1-v}   = \prod_{i=1}^N  \frac{1+v_i }{1-v_i} .
\end{equation}
Taking the logarithm on both sides and dividing the resulting expression by 2, we have
\begin{equation}
	\label{eqaddition4}
	\frac{1}{2} \ln\left(\frac{1+v}{1-v} \right)   = \frac{1}{2}  \sum_{i=1}^N  \ln \left(\frac{1+v_i }{1-v_i} \right).
\end{equation}
The rapidity $\sigma$ is defined as
\begin{equation}
	\label{eq13}
	\sigma\equiv \dfrac{1}{2}\ln\left(\dfrac{1+v/c}{1-v/c}\right)=\tanh^{-1}\left(\frac{v}{c}\right) \, , 
	\,\,\, \,\,\,\, \quad \sigma \in (-\infty, \infty) \, .
\end{equation}
Then Eq. \eqref{eqaddition4} is reduced to $\sigma=\sum_{i=1}^{N}\sigma_i$. This, means that the rapidity $\sigma$, which is a function of the relative velocity $v$,	satisfies the usual addition law of velocities and that the rapidity space is flat. As the rapidity is a function of relative velocities, it is clearly a Lorentz-invariant quantity.  On the other hand, in order to discuss the line element in the velocity space, and following Fock \cite{Fock64}, the relativistic relative velocity $v^\prime$ between two particles, which is a Lorentz-invariant quantity, is, in one dimension,
\begin{equation}
	\label{eq10}
	v^\prime=\dfrac{v_1 - v_2}{1-v_1v_2/c^2},
\end{equation}
where $v_1$ and $v_2$ are the velocities of the particles 1 and 2 as measured from the same inertial observer.

\par From \eqref{eq10}, and considering that both particles move with infinitesimally differing velocities, let us say, $v_1 = v$ and $v_2 = v + dv$, the square of the line element associated to the velocity space is 
\begin{equation}
	\label{eq11}
	ds^2\equiv\frac{d (v^\prime)^2}{c^2}=\frac{c^2 dv^2}{(c^2-v^2)^2}=\frac{\gamma^4}{c^2} dv^2  \,,
\end{equation}
where, clearly, $ds = d \sigma$. 
\par Considering now a relativistic gas of non-interacting particles whose rapidities (functions of the relative velocities of the particles with respect to the box) are independent variables, the Central Limit Theorem tell us that the sum $\sigma=\sum_{i=1}^{N}\sigma_i$ gives us, for a sufficiently large enough  number of particles the following distribution of the rapidities, or of the velocities, if we use  Eq. (\ref{eq13}),  
\begin{subequations}
	\begin{align}
	\label{Eq15}
	P_{LID}(\sigma)d\sigma &= \sqrt{\dfrac{\beta}{N\pi}} \, e^{-\frac{\beta\sigma^2}{N} } d\sigma \nonumber\\
	&= \sqrt{\dfrac{\beta}{N\pi c^2}} \,
	e^{-\frac{\beta  \tanh^{-1}(\frac{v}{c})^2}{N}} \gamma^2\left(v/c\right) dv \\
  \beta\equiv\dfrac{m_0c^2}{2kT}\,. 
\end{align}
\end{subequations}
As the rapidities are functions of the relative velocities of the particles with regard to the reference frame  (box), and as the correct one-dimensional pre-factor $\gamma^2$, required by the -- Lorentz-invariant -- Lobachevsky velocity space, appears naturally here, this distribution is a Lorentz-invariant distribution one (LID). Eq. \eqref{Eq15} recovers the Maxwell distribution of velocities for $v/c \ll 1$, and it does not admit velocities higher than the velocity of the light. Therefore, this Lorentz-invariant distribution is obtained by theoretical arguments, and its validity will be checked with numerical data obtained from dynamical molecular simulations. 

\par Then we can represent this Lorentz-invariant distribution as a function of the velocity but also as function of the rapidity, which is a more convenient variable to characterize a one-dimension relativistic gas, and whose domain is the whole set of real numbers, $\mathbb{R}$.  In fact, by representing the J\"uttner distribution in terms of the $\tanh^{-1}(v/c)$ it is possible to see that this distribution changes concavity at the origin for sufficiently high energy per particle, a behavior that is not observed if we use the Lorentz-invariant distribution (LID). As we can see in the next section, numerical simulations support the LID, showing no change of concavity at the origin. 

\par A key point about the invariance under a Lorentz transformation of a given PDF is that the temperature itself must be a Lorentz-invariant quantity, i.e., that the temperature measured by different inertial observers must be invariant. A direct consequence of that is the validity of  second law of thermodynamics  in special relativity. This fundamental point was raised by Landsberg in two seminal papers at the end of the 1960s \cite{Landsberg66,Landsberg67} and it is confirmed here.

\section{Mean Kinetic Energy} 
We evaluate the mean kinetic energy in one dimension
\begin{equation}
	E_{\text{dist}}=\int_{-\infty}^{\infty} P_{\text{dist}}(v) m_0 c^2 \gamma(v)dv \, ,
\end{equation}
where $E_{\text{dist}}$ and $P_{\text{dist}}$ are respectively the mean energy and the PDF of the distribution which is being used ($P_J(v)$, $P_{MJ}(v)$ or $P_{LID}(v)$). Thus,
for $P_{\text{dist}}(v)  = P_J(v)$, 
	\begin{equation}
		E_J= m_0c^2  \left(\dfrac{1}{\beta_{J}}+\dfrac{K_0(\beta_{J})}{K_1(\beta_J)}\right) \, ,
\end{equation}%
for $P_{\text{dist}}(v)  = P_{MJ}(v)$, 
\begin{equation}
E_{MJ} = \dfrac{m_0 c^2 K_1(\beta_{MJ})}{K_0(\beta_{MJ})} \, ,
\end{equation} 
for the LID $P_{\text{dist}}(v)  =P(v)$,
\begin{equation}
	E = m_0c^2 \exp\left(\dfrac{1}{4\beta }\right) \, ,
\end{equation}
and, for the Maxwell-Boltzmann distribution $P_{\text{dist}}(v)  = P_{MB}(v)$,
\begin{equation}
	E_{MB} =  \frac{k T_{MB}}{2 }   \, . 
\end{equation}

The present numerical results computed here were averaged over 200 relativistic molecular dynamical independent simulations. We considered a one-dimensional gas with $N=5000$ particles, half of them having a light mass, $m=1$, the other half having a heavy mass, $m=2$, with no interaction excepting perfect elastic binary collisions. Thus, we constructed histograms monitoring both the velocity $v$ and the rapidity variable $\sigma$.  

\par Our simulations correspond to a microcanonical ensemble, having a constant total energy $E$ in the box frame. We consider that,  in each binary collision,  the relativistic energy is conserved. To perform our simulations we solve the equations for the momentum and the energy conservation. Thus, before collision $(p_i,e_i)$ and after collision $(P_i,E_i)$, $i=1,2$, we have
\begin{align}
	P_1+P_2 &= p_1+p_2,\\
	E_1 + E_2 &= e_1 + e_2.
\end{align}
As a result, the momenta after the collision become 
\begin{align}
	P_i&=\left[\gamma(v_{zm})\right]^2\left[\dfrac{2v_{zm}e_i}{c^2} -\left(1+\dfrac{v_{zm}}{c}\right)^2p_i\right], 
\end{align}
where
\begin{equation}
	v_{zm}=\dfrac{(p_1+p_2)c^2}{e_1+e_2}
\end{equation}
is the velocity of the zero momentum frame, or the generalization of the center of mass in special relativity. With this, the time between two successive collisions is given by
\begin{equation}
	t=\dfrac{-v_{12} r_{12} - \sqrt{(v_{12}r_{12})^2-v_{12}^ 2r_{12}^2}}{v_{12}^2}
\end{equation}
where the relative position $r_{12}$ and velocity $v_{12}$ are respectively given by $r_{12}=r_1-r_2$ and $ v_{12}=(v_1-v_2)/(1-v_1v_2/c^2)$. Time $t$ is used to evolve the system. We consider the gas inside a one-dimensional box of length $L$ with periodic boundary conditions.      

\par Our results are in complete agreement with those obtained by Cubero et al \cite{cubero2007} for distribution at low temperatures $T\le 0.4$. We have numerically checked this by constructing histograms for the velocities of the particles with respect to the box, and then observing this system from external inertial frames  moving at different velocities with respect to the box. To compute the probability density function (PDF) for the rapidity, we arbitrarily choose, once for ever, a particle and measure the rapidities of all the others with respect to this one. The rapidity was computed directly from the velocities of the particles with respect to the laboratory frame, see Eqs. \eqref{eqaddition} and \eqref{eq13}. The rapidity histograms are constructed with constant rapidity bins\cite{CGS16}.
To minimize statistical fluctuations, we average $200$ simulations for a given total energy $E$.

Just for completeness, we have also performed these numerical simulations by using as inertial frame either the box or an arbitrarily chosen particle, and we have obtained the same  statistical distribution. 
The important point here is that we are using the rapidity of the relativistic difference of velocities, which is  a Lorentz-invariant 
quantity. The fact that these various ways to calculate the numerical distribution yield the same histogram  confirms the correctness of our simulations. Also, our numerical results for the velocities recover those of Cubero et al \cite{cubero2007}  for not very high temperatures.

\begin{figure}[h!]
	\begin{center}
	\begin{tabular}{cc}
		\includegraphics[width=7.5cm]{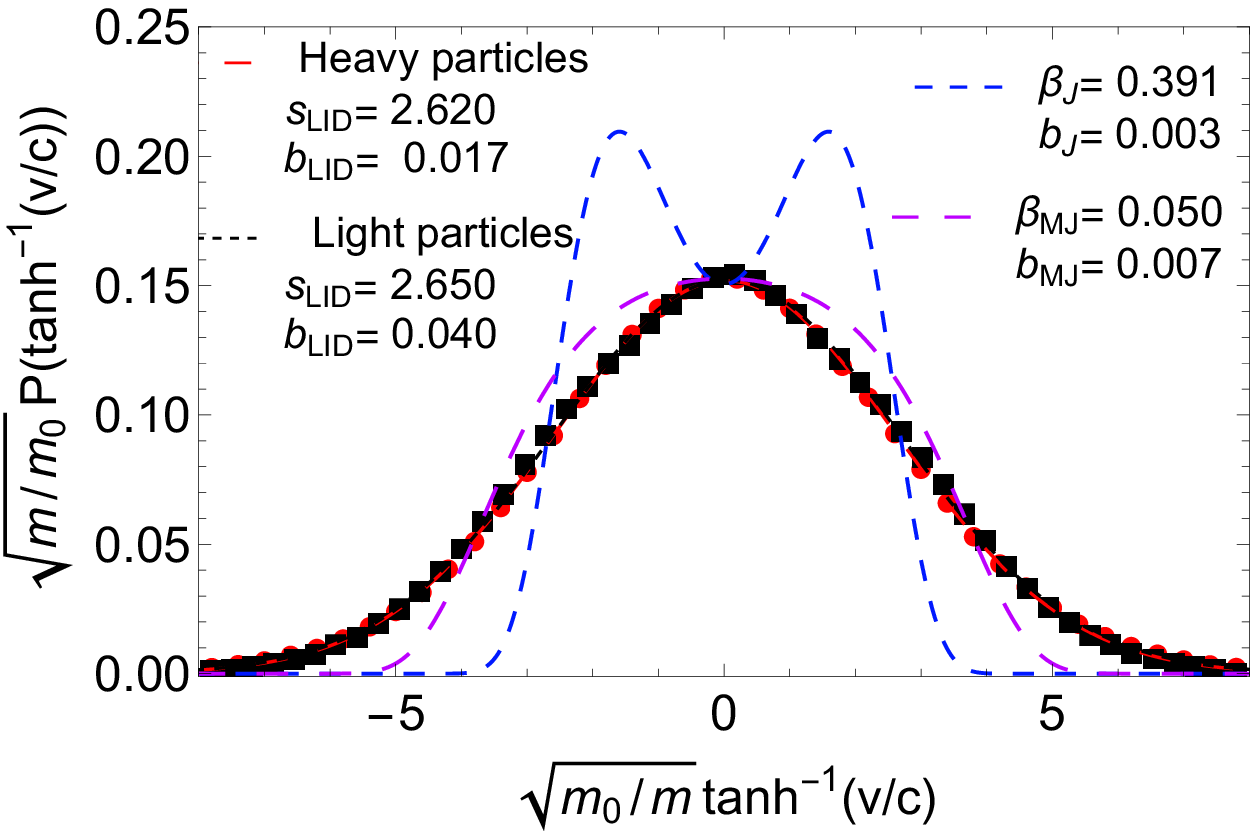} &
		 \includegraphics[width=7.5cm]{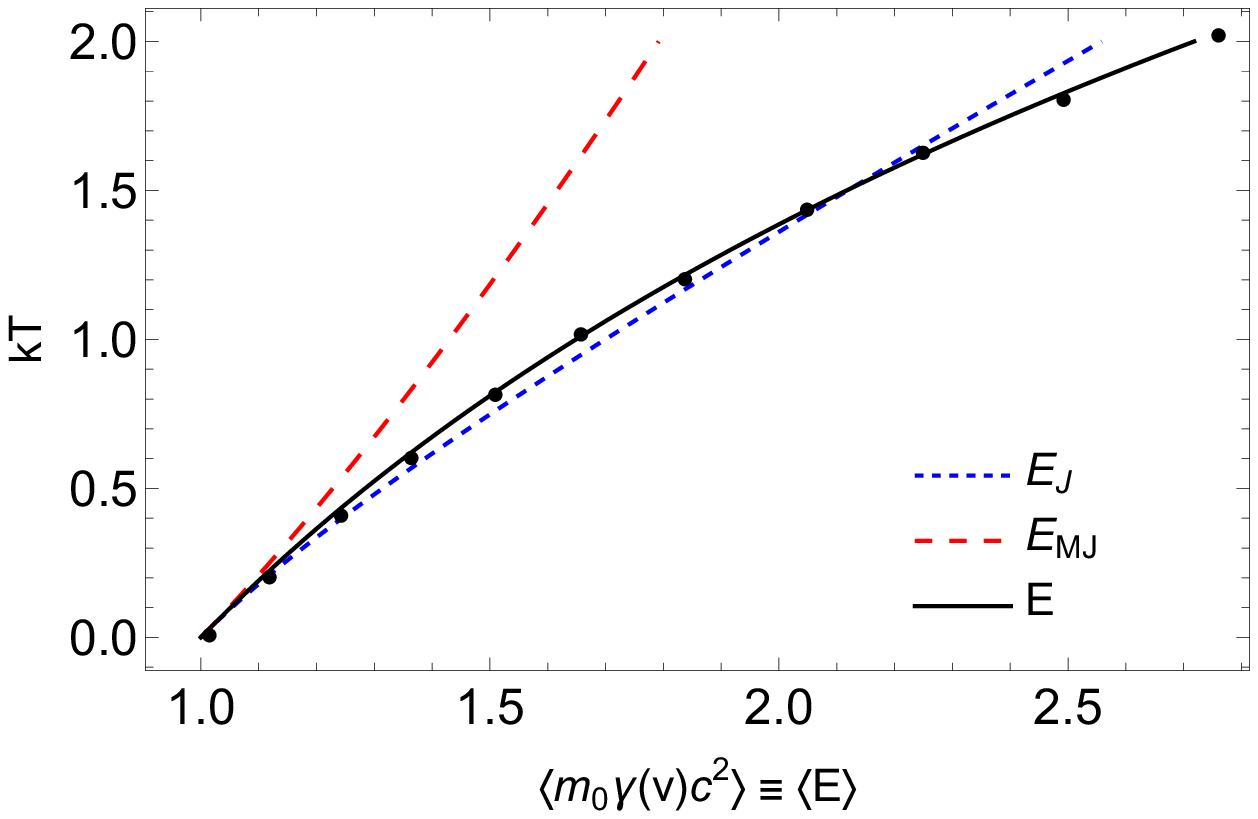}  
	\end{tabular}
	\end{center}
	\caption{Left panel: Numerical results of the simulations. Black and red points are the numerical results for heavy, $m_H$, and light, $m_L$, particles, respectively. Dashed red and black lines are the adjustments using $P_{LID}(\sigma)$, blue and purple dashed lines represent the J\"uttner  and the Modified J\"uttner  distributions respectively. Note that all three distributions asymptotically coincide only in the $v/c \rightarrow 0$ limit. Right panel: Temperature $T$ as a function of the mean energy $ \langle m_0 \gamma(v) c^2\rangle  \equiv \langle E\rangle $ for $P_{J}(v)$, $P_{MJ}(v)$ and $P_{LID}(v)$. Dashed blue, dashed red and continuous black represent $E_J$, $E_{MJ}$ and the present $E$ respectively. Black points correspond to numerical simulations for typical temperature values. All  three analytical curves recover, at low temperatures, the Maxwell-Botzmann behavior: $\langle E \rangle-m_0c^2 \sim k T/2\;\;(T \to 0)$.		
	}\label{fig7}
\end{figure}

In the left panel of figure \ref{fig1}, we show the numerical result of our simulations. The blue and the purple points correspond to the numerical data. We plot the Jüttner distribution, in particular for $T_J=0.18$ and for $T_J=1.01$, and compare with the blue and purple point distributions, respectively. We found that for lower values of $T_J\le 0.4$ Jüttner's distribution fits very well the distribution of points, however for values of $T_J$ greater that $0.4$, the differences between the point distribution and the Jüttner fitting grows in particular near to $v=\pm 1$. On the other hand, in the left panel of Fig. \ref{fig7} we show our numerical results in rapidity. The black and red points correspond to the numerical data, whereas the red and black lines are LID adjustments for the heavy and light particles respectively. Also, for comparison, we plot the J\"uttner  and the Modified J\"uttner distributions, exhibiting that the latter two can not adjust the numerical data. In variance, the numerical simulations support very well the LID. In the right panel of Fig. \ref{fig7}. we show in dashed blue, dashed red and continuous black the temperature, $T$, as a function of $E_{J}$, $E_{MJ}$ and $E$ respectively. Black points correspond to numerical simulations for different temperature values. $T_{MJ}> T_{J}$ and $T_{MJ} > T$ for all values of $\langle E \rangle$ . For $\langle m_0 c^2\gamma(v) \rangle <2.127$, $T> T_J$ where $\langle \rangle$ denotes average over velocities. Note that for $\langle m_0 c^2\gamma(v) \rangle >2.127$,  $T < T_J$. If $\langle m_0 c^2\gamma(v) \rangle =2.127$, $T_J=T=1.509$.    
\section{LID derivation from a MaxEnt principle }
Given that the rapidity is a Lorentz-invariant quantity, we could redefine $\tilde \sigma$ by convenience as $\tilde{\sigma}=\sigma/N$, then the entropy
\begin{equation}
	\label{eq101}
	S= -k \int_\infty^\infty\, P(\tilde\sigma) \ln P(\tilde\sigma) d \tilde\sigma,
\end{equation}
can be maximized with the constraints
\begin{equation}
	\label{eq102}
	\int_\infty^\infty\, P(\tilde\sigma) d \tilde\sigma = 1\,,\qquad \int_\infty^\infty\, \tilde\sigma^2 P(\tilde\sigma) d\tilde\sigma = \text{constant}\,.
\end{equation}
This yields
\begin{subequations}
\begin{align}
	& P(\tilde\sigma) d\tilde\sigma \equiv  P_{LID}(\sigma)d\sigma=\sqrt{\dfrac{\beta}{N\pi}} \, e^{-\frac{\beta\sigma^2}{N} } d\sigma \,, 
	\\
	& \beta\equiv \dfrac{m_0c^2}{2kT}\,.  
\end{align}
\end{subequations}
This distribution is a Lorentz-invariant one (LID).
\section{Conclusions}
We have discussed the original paper by J\"uttner (1911) and we have shown that his distribution, as well as a particular modification of the J\"uttner distribution, are not Lorentz-invariant distributions and, therefore, are not the correct relativistic generalizations of the Maxwell distribution of velocities of a gas. Essentially, the space of velocities considered by J\"uttner is a flat space while the correct relativistic velocity space is a Lobachevsky space.
Based on the relativistic addition of velocities and on the Central Limit Theorem, we have derived a truly Lorentz-invariant distribution (LID) that recovers the Maxwell distribution for $|v|/c \ll 1$ and has zero probability to have particles with speed higher than the speed of light. We have performed relativistic molecular dynamics simulations and our  numerical results confirm the LID for all values of $E$, whereas they are in odd with both the J\"uttner and the so-called Modified J\"uttner distribution at sufficiently high energies per particle of the gas.  It is worth to notice that all distributions $P_J(v)$, $P_{MJ}(v)$, $P_{LID}(v)$ and $P_{MB}(v)$ yield relations between the mean energies and the associated temperatures, some of which are shown in Fig. \ref{fig7}. Finally, an important consequence of the LID is that the temperature is, as argued by Landsberg, a Lorentz-invariant quantity, a crucial requirement to preserve the second law of thermodynamics in special relativity.

We are working in extensions of these one-dimensional results for two and three-dimensions. In these higher dimensions 
we have to consider the finite size of the particles during collisions, in molecular dynamics simulations. Also, the addition rules for the 
velocities are much more complicated and  there is no simple rule like the addition of rapidities in one dimension. But, certainly, mainly 
in three-dimensions, finding a correct Lorentz-invariant distribution is very important, including for the correct determination of the temperature of stellar objects.

\section*{acknowledgments}
	We acknowledge partial financial support by Conselho Nacional de Desenvolvimento Científico e Tecnológico (CNPq), Fundação Carlos Chagas Filho de Amparo à Pesquisa do Estado do Rio de Janeiro (FAPERJ), and Coordenação de Aperfeiçoamento de Pessoal de Nível Superior (CAPES) (Brazilian agencies).

\nocite{*}

\begin{thebibliography}{20}%
	\makeatletter
	\providecommand \@ifxundefined [1]{%
		\@ifx{#1\undefined}
	}%
	\providecommand \@ifnum [1]{%
		\ifnum #1\expandafter \@firstoftwo
		\else \expandafter \@secondoftwo
		\fi
	}%
	\providecommand \@ifx [1]{%
		\ifx #1\expandafter \@firstoftwo
		\else \expandafter \@secondoftwo
		\fi
	}%
	\providecommand \natexlab [1]{#1}%
	\providecommand \enquote  [1]{``#1''}%
	\providecommand \bibnamefont  [1]{#1}%
	\providecommand \bibfnamefont [1]{#1}%
	\providecommand \citenamefont [1]{#1}%
	\providecommand \href@noop [0]{\@secondoftwo}%
	\providecommand \href [0]{\begingroup \@sanitize@url \@href}%
	\providecommand \@href[1]{\@@startlink{#1}\@@href}%
	\providecommand \@@href[1]{\endgroup#1\@@endlink}%
	\providecommand \@sanitize@url [0]{\catcode `\\12\catcode `\$12\catcode
		`\&12\catcode `\#12\catcode `\^12\catcode `\_12\catcode `\%12\relax}%
	\providecommand \@@startlink[1]{}%
	\providecommand \@@endlink[0]{}%
	\providecommand \url  [0]{\begingroup\@sanitize@url \@url }%
	\providecommand \@url [1]{\endgroup\@href {#1}{\urlprefix }}%
	\providecommand \urlprefix  [0]{URL }%
	\providecommand \Eprint [0]{\href }%
	\providecommand \doibase [0]{http://dx.doi.org/}%
	\providecommand \selectlanguage [0]{\@gobble}%
	\providecommand \bibinfo  [0]{\@secondoftwo}%
	\providecommand \bibfield  [0]{\@secondoftwo}%
	\providecommand \translation [1]{[#1]}%
	\providecommand \BibitemOpen [0]{}%
	\providecommand \bibitemStop [0]{}%
	\providecommand \bibitemNoStop [0]{.\EOS\space}%
	\providecommand \EOS [0]{\spacefactor3000\relax}%
	\providecommand \BibitemShut  [1]{\csname bibitem#1\endcsname}%
	\let\auto@bib@innerbib\@empty
 
		\bibitem [1]{Sadegzadeh18}%
	\BibitemOpen
	\bibfield  {author} {\bibinfo {author} {\bibfnamefont {S.}~\bibnamefont
			{{Sadegzadeh}}}\ and\ \bibinfo {author} {\bibfnamefont {A.}~\bibnamefont
			{{Mousavi}}},\ }\href {\doibase 10.1063/1.5054830} {\bibfield  {journal}
		{\bibinfo  {journal} {Physics of Plasmas}\ }\textbf {\bibinfo {volume}
			{25}},\ \bibinfo {eid} {112107} (\bibinfo {year} {2018})}\BibitemShut
	{NoStop}%
 
		\bibitem [2]{Molnar20}%
	\BibitemOpen
	\bibfield  {author} {\bibinfo {author} {\bibfnamefont {S.~M.}\ \bibnamefont
			{{Molnar}}}\ and\ \bibinfo {author} {\bibfnamefont {J.}~\bibnamefont
			{{Godfrey}}},\ }\href {\doibase 10.3847/1538-4357/abb6f6} {\bibfield
		{journal} {\bibinfo  {journal} {Astrophysical Journal}\ }\textbf {\bibinfo
			{volume} {902}},\ \bibinfo {eid} {143} (\bibinfo {year} {2020})}\BibitemShut
	{NoStop}%
	\bibitem [3]{Melrose21}%
	\BibitemOpen
	\bibfield  {author} {\bibinfo {author} {\bibfnamefont {D.~B.}\ \bibnamefont
			{{Melrose}}}, \bibinfo {author} {\bibfnamefont {M.~Z.}\ \bibnamefont
			{{Rafat}}}, \ and\ \bibinfo {author} {\bibfnamefont {A.}~\bibnamefont
			{{Mastrano}}},\ }\href {\doibase 10.1093/mnras/staa3324} {\bibfield
		{journal} {\bibinfo  {journal} {MNRAS}\ }\textbf {\bibinfo {volume} {500}},\
		\bibinfo {pages} {4530} (\bibinfo {year} {2021})}\BibitemShut {NoStop}%
 
		\bibitem [4]{Adcox01}%
	\BibitemOpen
	\bibfield  {author} {\bibinfo {author} {\bibfnamefont {K.}~\bibnamefont
			{{Adcox}}}\ and\ \bibinfo {author} {\bibnamefont {{et al.}}} (\bibinfo
		{collaboration} {PHENIX Collaboration}),\ }\href {\doibase
		10.1103/PhysRevLett.88.022301} {\bibfield  {journal} {\bibinfo  {journal}
			{Phys. Rev. Lett.}\ }\textbf {\bibinfo {volume} {88}},\ \bibinfo {pages}
		{022301} (\bibinfo {year} {2001})}\BibitemShut {NoStop}%
	\bibitem [5]{Adler02}%
	\BibitemOpen
	\bibfield  {author} {\bibinfo {author} {\bibfnamefont {C.}~\bibnamefont
			{{Adler}}}\ and\ \bibinfo {author} {\bibnamefont {{et al.}}} (\bibinfo
		{collaboration} {STAR Collaboration}),\ }\href {\doibase
		10.1103/PhysRevLett.89.202301} {\bibfield  {journal} {\bibinfo  {journal}
			{Phys. Rev. Lett.}\ }\textbf {\bibinfo {volume} {89}},\ \bibinfo {pages}
		{202301} (\bibinfo {year} {2002})}\BibitemShut {NoStop}%
 
	\bibitem [6]{Marques15}%
	\BibitemOpen
	\bibfield  {author} {\bibinfo {author} {\bibfnamefont {L.}~\bibnamefont
			{{Marques}}}, \bibinfo {author} {\bibfnamefont {J.}~\bibnamefont
			{{Cleymans}}}, \ and\ \bibinfo {author} {\bibfnamefont {A.}~\bibnamefont
			{{Deppman}}},\ }\href {\doibase 10.1103/PhysRevD.91.054025} {\bibfield
		{journal} {\bibinfo  {journal} {Phys. Rev. D}\ }\textbf {\bibinfo {volume}
			{91}},\ \bibinfo {pages} {054025} (\bibinfo {year} {2015})}\BibitemShut
	{NoStop}%
 
	\bibitem [7]{Juttner}%
	\BibitemOpen
	\bibfield  {author} {\bibinfo {author} {\bibfnamefont {F.}~\bibnamefont
			{J\"uttner}},\ }\href{https://doi.org/10.1002/andp.19113390503}{\bibfield  {journal} {\bibinfo  {journal}
			{Annalen der Physik}\ }\textbf {\bibinfo {volume} {339}},\ \bibinfo {pages}
		{856} (\bibinfo {year} {1911})}\BibitemShut {NoStop}%
 
	\bibitem [8]{boltzmann1872}%
	\BibitemOpen
	\bibfield  {author} {\bibinfo {author} {\bibfnamefont {L.}~\bibnamefont
			{{Boltzmann}}},\ }\href{https://doi.org/10.1007/978-3-322-84986-1_3}{\bibfield  {journal} {\bibinfo  {journal}
			{Wiener Berichte}\ }\textbf {\bibinfo {volume} {66}},\ \bibinfo {pages} {275}
		(\bibinfo {year} {1872})}\BibitemShut {NoStop}%
 
	\bibitem [9]{boltzmann1877}%
	\BibitemOpen
	\bibfield  {author} {\bibinfo {author} {\bibfnamefont {L.}~\bibnamefont
			{{Boltzmann}}},\ }\href{https://doi.org/10.1017/CBO9781139381437.011}{\bibfield  {journal} {\bibinfo  {journal}
			{Wiener Berichte}\ }\textbf {\bibinfo {volume} {76}},\ \bibinfo {pages} {373}
		(\bibinfo {year} {1877})}\BibitemShut {NoStop}%
 
	\bibitem [10]{Maxwell_1}%
	\BibitemOpen
	\bibfield  {author} {\bibinfo {author} {\bibfnamefont {J.~C.}\ \bibnamefont
			{{Maxwell}}},\ }\href {\doibase 10.1080/14786446008642818} {\bibfield
		{journal} {\bibinfo  {journal} {The London, Edinburgh, and Dublin
				Philosophical Magazine and Journal of Science}\ }\textbf {\bibinfo {volume}
			{19}},\ \bibinfo {pages} {19} (\bibinfo {year}
		{1860}{\natexlab{a}})}\BibitemShut {NoStop}%
 
	\bibitem [11]{Maxwell_2}%
	\BibitemOpen
	\bibfield  {author} {\bibinfo {author} {\bibfnamefont {J.~C.}\ \bibnamefont
			{{Maxwell}}},\ }\href {\doibase 10.1080/14786446008642902} {\bibfield
		{journal} {\bibinfo  {journal} {The London, Edinburgh, and Dublin
				Philosophical Magazine and Journal of Science}\ }\textbf {\bibinfo {volume}
			{20}},\ \bibinfo {pages} {21} (\bibinfo {year}
		{1860}{\natexlab{b}})}\BibitemShut {NoStop}%
 
	\bibitem [12]{Landsberg66}%
	\BibitemOpen
	\bibfield  {author} {\bibinfo {author} {\bibfnamefont {P.~T.}\ \bibnamefont
			{{Landsberg}}},\ }\href {\doibase 10.1038/212571a0} {\bibfield  {journal}
		{\bibinfo  {journal} {Nature}\ }\textbf {\bibinfo {volume} {212}},\ \bibinfo
		{pages} {571} (\bibinfo {year} {1966})}\BibitemShut {NoStop}%
 
	\bibitem [13]{Landsberg67}%
	\BibitemOpen
	\bibfield  {author} {\bibinfo {author} {\bibfnamefont {P.~T.}\ \bibnamefont
			{{Landsberg}}},\ }\href {\doibase 10.1038/214903a0} {\bibfield  {journal}
		{\bibinfo  {journal} {Nature}\ }\textbf {\bibinfo {volume} {214}},\ \bibinfo
		{pages} {903} (\bibinfo {year} {1967})}\BibitemShut {NoStop}%
 
	\bibitem [14]{cubero2007}%
	\BibitemOpen
	\bibfield  {author} {\bibinfo {author} {\bibfnamefont {D.}~\bibnamefont
			{{Cubero}}}, \bibinfo {author} {\bibfnamefont {J.}~\bibnamefont
			{{Casado-Pascual}}}, \bibinfo {author} {\bibfnamefont {J.}~\bibnamefont
			{{Dunkel}}}, \bibinfo {author} {\bibfnamefont {P.}~\bibnamefont {{Talkner}}},
		\ and\ \bibinfo {author} {\bibfnamefont {P.}~\bibnamefont {{H{\"a}nggi}}},\
	}\href {\doibase 10.1103/PhysRevLett.99.170601} {\bibfield  {journal}
		{\bibinfo  {journal} {Phys. Rev. Lett.}\ }\textbf {\bibinfo {volume} {99}},\
		\bibinfo {eid} {170601} (\bibinfo {year} {2007})}\BibitemShut {NoStop}%
	\bibitem [15]{Montakhab09}%
	\BibitemOpen
	\bibfield  {author} {\bibinfo {author} {\bibfnamefont {A.}~\bibnamefont
			{{Montakhab}}}, \bibinfo {author} {\bibfnamefont {M.}~\bibnamefont
			{{Ghodrat}}}, \ and\ \bibinfo {author} {\bibfnamefont {M.}~\bibnamefont
			{{Barati}}},\ }\href {\doibase 10.1103/PhysRevE.79.031124} {\bibfield
		{journal} {\bibinfo  {journal} {Phys. Rev. E}\ }\textbf {\bibinfo {volume}
			{79}},\ \bibinfo {pages} {031124} (\bibinfo {year} {2009})}\BibitemShut
	{NoStop}%
	\bibitem [16]{Ghodrat11}%
	\BibitemOpen
	\bibfield  {author} {\bibinfo {author} {\bibfnamefont {M.}~\bibnamefont
			{{Ghodrat}}}\ and\ \bibinfo {author} {\bibfnamefont {A.}~\bibnamefont
			{{Montakhab}}},\ }\href {\doibase https://doi.org/10.1016/j.cpc.2011.01.018}
	{\bibfield  {journal} {\bibinfo  {journal} {Computer Physics Communications}\
		}\textbf {\bibinfo {volume} {182}},\ \bibinfo {pages} {1909} (\bibinfo {year}
		{2011})},\ \bibinfo {note} {computer Physics Communications Special Edition
		for Conference on Computational Physics Trondheim, Norway, June 23-26,
		2010}\BibitemShut {NoStop}%
	\bibitem [17]{CGS16}%
	\BibitemOpen
	\bibfield  {author} {\bibinfo {author} {\bibfnamefont {E. M. F.}~\bibnamefont
			{{Curado}}}, \bibinfo {author} {\bibfnamefont {F. T. L.}~\bibnamefont
			{{Germani}}} and\ \bibinfo {author} {\bibfnamefont {I. D.}~\bibnamefont
			{{Soares}}},\ }\href {\doibase 10.1063/1.2165771} {\bibfield  {journal}
		{\bibinfo  {journal} {Physica A: Statistical Mechanics and its Applications}\ }\textbf {\bibinfo
			{volume} {444}},\ \bibinfo {pages} {963-969} (\bibinfo {year}
		{2016})}\BibitemShut {NoStop}%
	\bibitem [18]{Kubli2021}%
\BibitemOpen
\bibfield  {author} {\bibinfo {author} {\bibfnamefont {N.}~\bibnamefont
		{{Kubli}}}\ and\ \bibinfo {author} {\bibfnamefont {H.~J.}\ \bibnamefont
		{{Herrmann}}},\ }\href {\doibase 10.1016/j.physa.2020.125261} {\bibfield
	{journal} {\bibinfo  {journal} {Physica A Statistical Mechanics and its
			Applications}\ }\textbf {\bibinfo {volume} {561}},\ \bibinfo {eid} {125261}
	(\bibinfo {year} {2021})}\BibitemShut {NoStop}%
	\bibitem [19]{Lehmann06}%
	\BibitemOpen
	\bibfield  {author} {\bibinfo {author} {\bibfnamefont {E.}~\bibnamefont
			{{Lehmann}}},\ }\href {\doibase 10.1063/1.2165771} {\bibfield  {journal}
		{\bibinfo  {journal} {Journal of Mathematical Physics}\ }\textbf {\bibinfo
			{volume} {47}},\ \bibinfo {pages} {023303} (\bibinfo {year}
		{2006})}\BibitemShut {NoStop}%
	\bibitem [20]{Dunkel_Hanggi_07}%
	\BibitemOpen
	\bibfield  {author} {\bibinfo {author} {\bibfnamefont {J.}~\bibnamefont
			{{Dunkel}}}\ and\ \bibinfo {author} {\bibfnamefont {P.}~\bibnamefont
			{{H\"anggi}}},\ }\href {\doibase https://doi.org/10.1016/j.physa.2006.07.013}
	{\bibfield  {journal} {\bibinfo  {journal} {Physica A: Statistical Mechanics
				and its Applications}\ }\textbf {\bibinfo {volume} {374}},\ \bibinfo {pages}
		{559} (\bibinfo {year} {2007})}\BibitemShut {NoStop}%
	\bibitem [21]{Fock64}%
	\BibitemOpen
	\bibfield  {author} {\bibinfo {author} {\bibfnamefont {V.}~\bibnamefont
			{{Fock}}},\ }\href {\doibase 10.1016/B978-0-08-010061-6.50008-1} {\emph
		{\bibinfo {title} {The Theory of Space, Time and Gravitation}}},\ \bibinfo
	{edition} {2nd}\ ed.\ (\bibinfo  {publisher} {Pergamon Press},\ \bibinfo
	{year} {1964})\ Chap.\ \bibinfo {chapter} {{I} - the theory of relativity},
	pp.\ \bibinfo {pages} {9--53}\BibitemShut {NoStop}%
\end{thebibliography}
\providecommand{\noopsort}[1]{}\providecommand{\singleletter}[1]{#1}%

\end{document}